\documentclass[a4paper,12pt, epsfig]{article}
\usepackage{epsfig}
\usepackage{amssymb}
\usepackage{amsfonts}
\usepackage{amsmath}

\newskip\humongous \humongous=0pt plus 1000pt minus 1000pt

\newif\ifdtup

\catcode`\@=11

\@addtoreset{equation}{section}
\def\theequation{\thesection.\arabic{equation}}

\def\@normalsize{\@setsize\normalsize{15pt}\xiipt\@xiipt
\abovedisplayskip 14pt plus3pt minus3pt%
\belowdisplayskip \abovedisplayskip
\abovedisplayshortskip \z@ plus3pt%
\belowdisplayshortskip 7pt plus3.5pt minus0pt}

\def\small{\@setsize\small{13.6pt}\xipt\@xipt
\abovedisplayskip 13pt plus3pt minus3pt%
\belowdisplayskip \abovedisplayskip
\abovedisplayshortskip \z@ plus3pt%
\belowdisplayshortskip 7pt plus3.5pt minus0pt
\def\@listi{\parsep 4.5pt plus 2pt minus 1pt
     \itemsep \parsep
     \topsep 9pt plus 3pt minus 3pt}}

\relax

\catcode`@=12

\catcode`\@=11

\def\section{\@startsection{section}{1}{\z@}{3.5ex plus 1ex minus
   .2ex}{2.3ex plus .2ex}{\large\bf}}

\def\thesection{\arabic{section}}
\def\thesubsection{\arabic{section}.\arabic{subsection}}

\def\appendix{\setcounter{section}{0}
 \def\thesection{Appendix \Alph{section}}
 \def\thesubsection{\Alph{section}.\arabic{subsection}}
 \def\theequation{\Alph{section}.\arabic{equation}}}

\def\SymBoxes#1#2#3#4{\newdimen\un@t \un@t#3%
\raisebox{#1}{\rule{#2\un@t}{#4}\hskip-#2\un@t
\@tempdimb\un@t \advance\@tempdimb by-#4\@tempcntb#2\relax%
\@whilenum{\@tempcntb>0}\do{
\rule{#4}{\un@t}\hskip\@tempdimb \advance\@tempcntb by\m@ne}%
\hskip-#2\un@t \rule[\un@t]{#2\un@t}{#4}%
\rule[\un@t]{#4}{#4}\hskip-#4
\rule{#4}{\un@t}}\hskip-#4}                

\begin{document}

\newcommand{\beq}{\begin{equation}}
\newcommand{\eeq}{\end{equation}}
\newcommand{\bea}{\begin{eqnarray}}
\newcommand{\eea}{\end{eqnarray}}
\newcommand{\beas}{\begin{eqnarray*}}
\newcommand{\eeas}{\end{eqnarray*}}
\newcommand{\defi}{\stackrel{\rm def}{=}}
\newcommand{\non}{\nonumber}
\newcommand{\bquo}{\begin{quote}}
\newcommand{\enqu}{\end{quote}}
\renewcommand{\(}{\begin{equation}}
\renewcommand{\)}{\end{equation}}
\def\IZ{{\mathbb Z}}
\def\IR{{\mathbb R}}
\def\IC{{\mathbb C}}
\def\IQ{{\mathbb Q}}

\def\CM{{\mathcal{M}}} 
\def\dCM{{\left \vert\mathcal{M}\right\vert}}

\def \d{\textrm{d}}
\def \p{\partial}
\def\Tr{ \hbox{\rm Tr}}
\def\half{\frac{1}{2}}

\def \eqn#1#2{\begin{equation}#2\label{#1}\end{equation}}
\def\de{\partial}
\def\Tr{ \hbox{\rm Tr}}
\def\H{ \hbox{\rm H}}
\def\HE{ \hbox{$\rm H^{even}$}}
\def\HO{ \hbox{$\rm H^{odd}$}}
\def\K{ \hbox{\rm K}}
\def\Im{ \hbox{\rm Im}}
\def\Ker{ \hbox{\rm Ker}}
\def\const{\hbox {\rm const.}}
\def\o{\over}
\def\im{\hbox{\rm Im}}
\def\re{\hbox{\rm Re}}
\def\bra{\langle}\def\ket{\rangle}
\def\Arg{\hbox {\rm Arg}}
\def\Re{\hbox {\rm Re}}
\def\Im{\hbox {\rm Im}}
\def\exo{\hbox {\rm exp}}
\def\diag{\hbox{\rm diag}}
\def\longvert{{\rule[-2mm]{0.1mm}{7mm}}\,}
\def\a{\alpha}
\def\dag{{}^{\dagger}}
\def\tq{{\widetilde q}}
\def\p{{}^{\prime}}
\def\W{W}
\def\N{{\cal N}}
\def\hsp{,\hspace{.7cm}}
\newcommand{\C}{\ensuremath{\mathbb C}}
\newcommand{\Z}{\ensuremath{\mathbb Z}}
\newcommand{\R}{\ensuremath{\mathbb R}}
\newcommand{\rp}{\ensuremath{\mathbb {RP}}}
\newcommand{\cp}{\ensuremath{\mathbb {CP}}}
\newcommand{\vac}{\ensuremath{|0\rangle}}
\newcommand{\vact}{\ensuremath{|00\rangle}}
\newcommand{\oc}{\ensuremath{\overline{c}}}

\begin{titlepage}
\begin{flushright}
ULB-TH/08-06\\
\end{flushright}
\bigskip
\def\thefootnote{\fnsymbol{footnote}}

\begin{center}
{\Large {\bf The Mesonic Branch of the Deformed Conifold \\}}
\end{center}

\bigskip
\begin{center}
{\large  Chethan 
KRISHNAN\footnote{\texttt{Chethan.Krishnan@ulb.ac.be}} and Stanislav 
KUPERSTEIN\footnote{\texttt{skuperst@ulb.ac.be}}}\\
\end{center}

\renewcommand{\thefootnote}{\arabic{footnote}}

\begin{center}
\vspace{1em}
{\em  { International Solvay Institutes,\\
Physique Th\'eorique et Math\'ematique,\\
ULB C.P. 231, Universit\'e Libre
de Bruxelles, \\ B-1050, Bruxelles, Belgium\\}}

\end{center}

\noindent
\begin{center} {\bf Abstract} \end{center}
Using coordinates that manifest the $S^2$-$S^3$ split of the base, we 
study D3-branes localized 
on the three-sphere in the Klebanov-Strassler background.
We find a numerical solution for the warp factor and show the emergence of 
the 
AdS throat near the stack.
In the dual gauge theory, this corresponds to an RG flow along the 
mesonic branch.
We demonstrate how the cubic superpotential of the $\mathcal{N}=4$ SYM
theory emerges at the end of the RG flow.

\vspace{1.6 cm}

\vfill

\end{titlepage}
\bigskip

\hfill{}
\bigskip

\tableofcontents

\setcounter{footnote}{0}
\section{\bf Introduction}

\noindent

The AdS/CFT correspondence \cite{Maldacena} enunciates that the low energy 
effective 4d physics on a heavy stack of $D3$-branes at a smooth point in 
flat 
space-time is dual to the 
near horizon limit of the 10d curved (by the $D$-branes) geometry. 
More specifically, the 
original AdS/CFT conjecture 
proposed a duality between $\mathcal{N}=4$
$SU(N)$ SYM gauge theory and type IIB supergravity on $AdS_5 
\times S^5$. 

One powerful way of constructing gauge theories with less supersymmetry 
is to consider instead stacks of $D3$-branes at the singular tip of a Calabi-Yau 
cone $X_6$ \cite{Klebanov:1998hh, Morrison}. Exactly like in the flat 
space case the radial coordinate
of $X_6$ is absorbed in the $AdS_5$ part of the metric and the near horizon geometry
becomes $AdS_5 \times Y^5$, where $Y^5$ is the $5d$  base of $X_6$. The 
number of supercharges 
in the dual gauge theory is completely encoded in $Y^5$ and is maximal 
only for $Y^5=S^5$.

The most notorious  and well studied $\mathcal{N}=1$ example of this 
kind is the Klebanov-Witten model,
which arises from $D3$-branes placed at the tip of the singular conifold. It was argued in \cite{Klebanov:1998hh}
that the low energy effective gauge theory living on the stack has a non-trivial RG fixed point and, therefore,
is conformal. The gauge group is $SU(N) \times SU(N)$ and the field content consists of four chiral bi-fundamentals $A_{1,2}$ and $B_{1,2}$  that transform in the $(\mathbf{N},\bar{\mathbf{N}})$ 
and the $(\mathbf{N},\bar{\mathbf{N}})$ representations respectively. The theory has also a marginal superpotential $W \propto \Tr \det_{ij} A_i B_j$. The $5d$ base of the conifold $T^{1,1}$ 
is topologically $S^3 \times S^2$ and has an 
$SU(2) \times SU(2) \times U(1)_R$ isometry, which appears also as the global symmetry of the gauge theory.
The two $SU(2)$ factors act on $A_i$'s and $B_i$'s respectively and the non-trivial R-symmetry charges are $\frac{1}{2}$ for all the fields. It is straightforward to check that with this assignment the gauge couplings do not run. The theory enjoys also an additional non-geometric baryonic symmetry $U(1)_B$  and a chiral
$\mathbb{Z}_2$ symmetry, which interchanges $A_i$'s and $B_i$'s and also the two $SU(2)$ groups.

The conformal properties of the gauge theory are encoded in the $AdS_5$ part of the metric.
This is evident from the fact that the $4d$ conformal group is isomorphic to the $AdS_5$ isometry group $SO(4,2)$.
As we have already mentioned, the $AdS_5$ factor owes its appearance to 
the conic structure of the $6d$ CY
space $X_6$.
It follows therefore, that in order to build a non-conformal extension of 
the 
AdS/CFT duality (the so-called 
non-AdS/non-CFT correspondence \cite{AharonyR, F}) we have to 
change the conic structure 
of $X_6$, while still possibly keeping some of the supersymmetries. 
For the conifold there are two ways to achieve this goal. The 
\emph{deformation} changes the complex structure of the conifold, but 
still keeps the K\"ahler structure, while the \emph{resolution} of the 
conifold breaks the K\"ahler but preserves the complex structure. Though 
both the deformation and the resolution make the conifold
completely regular and smooth, they look different at the tip. In the 
former case the $S^3$ of $T^{1,1}$
approaches a finite size and the $S^2$ shrinks to zero, while in the latter case the situation is exactly the opposite.

The supersymmetric supergravity solution based on the deformed conifold 
was constructed by Klebanov and Strassler \cite{KS}
and has since been a subject of intensive research. The solution necessarily incorporates
$M$ fractional $D3$-branes, which are actually regular $D5$-branes wrapped on the two-sphere. On the gauge
theory side it means that the gauge group is now $SU\left((k+1)M \right) \times SU(kM)$. The theory, as expected, is not conformal.
When one gauge group becomes weakly coupled, the
other becomes strongly coupled. Under Seiberg duality, however, the r\^oles of the couplings are exchanged,
while the gauge group  becomes $SU(kM) \times SU\left((k-1)M \right)$.  
The theory exhibits, therefore, a \emph{cascade} of Seiberg dualities. At 
each step of the cascade we have $k \to k-1$ and at the last step we 
arrive at the $SU(M)$ $\mathcal{N}=1$ SYM. It was first suggested by 
Aharony \cite{Aharony} that the theory is at a specific 
$\mathbb{Z}_2$-invariant
point on the \emph{baryonic} branch $\vert A \vert = \vert B \vert$. 
The broken baryonic symmetry $U(1)_B$ thus implies that the gauge theory 
has a pseudoscalar Goldstone boson and its massless scalar superpartner. 
The supergravity dual of these modes was later found in \cite{Gubser}.
The baryonic branch allows also for solutions that break the   
$\mathbb{Z}_2$ symmetry \cite{DKS}.
The corresponding supergravity duals based on the so-called resolved warped deformed conifold
were constructed in \cite{Butti} (see also \cite{DKS, Riccardo}).

In this paper we want to construct a gravity dual of the \emph{mesonic} branch of the gauge theory.
In this case the gauge group is $SU(\widetilde{N}+M) \times SU(\widetilde{N})$ and the cascade step is simply given by $\widetilde{N} \to \widetilde{N}-M$.  When $\widetilde{N}$ becomes smaller than $M$ 
no Seiberg dual description  exists anymore. Instead the superpotential receives a non-perturbative
Affleck-Dine-Seiberg (ADS) contribution and the quantum moduli space describes $M$ copies of the deformed conifold.
The $SU(\widetilde{N}+M)$ gauge group is broken by the meson VEVs, while the deformation parameter of
the conifold depends on the strong coupling scale of the surviving $SU(M)$ gauge group.
The branch essentially describes $\widetilde{N}$ $D3$-branes moving on the deformed conifold. 
Actually, a similar branch exists also for $\widetilde{N}>M$. It 
corresponds to mesons acquiring large enough VEVs,
such that the quantum corrections can be captured by an ADS like 
term in the superpotential.  We will now have $N$ $D3$-branes moving on the deformed conifold, where $N$
is the value of $\widetilde{N}$ at the specific step of the cascade.

On the supergravity side the setup should include both $M$ fractional branes of the original KS background
and $N$ physical $D3$-branes. To produce a regular $10d$ solution we have to localize the $D3$-branes at a point
on the conifold. We will be interested in a $D3$-brane stack placed at the ``tip", namely the North pole
of the blown-up three sphere. 
Back in the gauge theory, this corresponds to $N$ mesons receiving the 
\emph{same} VEVs. The RG flow triggered by the VEV will end in the 
$\mathcal{N}=4$
$SU(N)$ SYM gauge theory, just because we put the $D3$-branes at a regular point.
The backreaction of the brane stack yields the $AdS_5$ throat, so the 
entire supergravity solution describes the flow\footnote
{To be more precise the KS solution reproduces the $AdS_5 \times T^{1,1}$
geometry in the UV only up to logarithmic corrections, which just indicates the fact that in the KS model 
the gauge theory in the UV is not $SU(N) \times SU(N)$. }
from $AdS_5 \times T^{1,1}$
to $AdS_5 \times S^5$.

Our approach is partially based on the work of Klebanov and Murugan 
\cite{Klebanov:2007us}. (See \cite{Cvetic:2007nv, Martelli:2007mk, KK} 
also for closely related work.). They studied 
a similar emergence of the $AdS_5$ throat due to a $D3$-brane stack 
located at a point on the
blown-up two sphere of the resolved conifold\footnote
{The singular solution corresponding to $D3$-branes smeared on the $S^2$ was
investigated in \cite{Pando}.}.
On the gauge theory side this describes an RG flow along the 
\emph{non-mesonic} branch.

Finding the full 10d solution in \cite{Klebanov:2007us} was equivalent to 
solving the 6d Laplace equation with a source for the warp function. 
In our case, we want to add source D3-branes instead to the 
Klebanov-Strassler 
background, which is more complicated because of the extra fields etc.
But when we add sources, the 6d inhomogeneous Laplace 
equation is still the only equation we need to consider, because 
we are still working within the framework of the standard D3-brane ansatz.
Indeed, the KS warp function satisfies:
\beq
\Box_{6} h_{\textrm{KS}} = g_s \star_6 H^{\textrm{KS}}_3 \wedge F^{\textrm{KS}}_3,
\eeq
where $ H^{\textrm{KS}}_3$ and $F^{\textrm{KS}}_3$ are the NS-NS and RR $3$-forms.
There is no source term on the right hand side of the equation, which shows that there are 
no $D3$-branes in the background, but rather only $M$ fractional branes. 
On the other hand, the corresponding RR charge $\widetilde{N} $ is non zero and the asymptotic behavior of the 
self-dual RR-form is:
\beq
\widetilde{F}_5 \approx \widetilde{N} \, \textrm{Vol} \left( T^{1,1} \right),
\qquad \textrm{where} \qquad
\widetilde{N} = \frac{3}{2 \pi } g_s \ln \frac{r}{r_0} \cdot M.
\eeq
To build our background we have to split the warp function into two terms:
\beq
h = h_{\textrm{KS}} + H_{\textrm{D3}},
\eeq
where $H_{\textrm{D3}}$ (or simply $H$ throughtout the paper)
is the solution of the Laplace equation with the $D3$-brane source:
\beq
\Box_{6} H_{\textrm{D3}} = N \delta_{6}(\textrm{NP}),
\eeq 
where $\textrm{NP}$ stands for the North pole of the $S^3$. Now we have:
\beq
\label{eq:tildeN}
\widetilde{N} = N+\frac{3}{2 \pi } g_s \ln \frac{r}{r_0} \cdot M.
\eeq
It is essential to notice that the addition of the source term is 
consistent with the usual ansatz for D3-branes. In particular, the dilaton 
is constant, the 0-form vanishes and SUSY is not broken.

The organization of the paper is as follows.
In the next section and in one of the appendices, we describe the 
deformed conifold. 
We introduce a new map which for a given point on the deformed conifold provides
its $S^3$ and $S^2$ coordinates. The map generalizes the results of \cite{EK}
for the singular conifold case.
We then relate this map to the coordinates introduced in \cite{Gimon}
and later used in \cite{Sakai}.
These coordinates are different from the standard coordinates used by  
Klebanov and Strassler \cite{Minasian, KS}, and
prove to be very convenient for working with the Laplace equation. We  
explain this in detail in Section \ref{LE}, where we present the numeric 
solution of the equation
and demonstrate the emergence of the AdS throat.
Section \ref{GT} is devoted to the gauge theory. We show how the cubic superpotential
of the $\mathcal{N}=4$ theory emerges when one expands the mesonic fields
around the VEV corresponding to the North pole of the three-sphere.
We end with some remarks in Section \ref{D}. In particular we propose why 
the 
gravity mode \cite{Gubser} dual to the Goldstone boson of the baryonic 
symmetry does not exist in our case.
Some of the technicalities have been relegated to various appendices.

\section{\bf The Deformed Conifold}

\label{DCP}

We start with a brief description of the (singular) conifold. In the 
physics 
community\footnote{In mathematics, the notion of a conifold is
more general. It refers to a generalization of the notion of a manifold,
where we allow conical singularities. The physics-conifold is a special 
case of the mathematics-conifold.} the word conifold
refers to 
the singular non-compact Calabi-Yau three-fold defined 
by the complex quadratic equation
\eqn{conifold}{\sum_{i=1}^{4}z_i^2=0.}
This equation represents a real cone over a five-dimensional 
Einstein manifold called $T^{1,1}$, which is the coset space 
$(SU(2)\times SU(2))/U(1)$. The base $T^{1,1}$ has the 
topology of $S^2\times S^3$ \cite{Candelas:1989js}, and if we denote the 
metric on it by $d\Omega^2_{T^{1,1}}$ then the full conifold metric takes 
the standard form of a cone: $ds_6^2=dr^2+r^2d\Omega^2_{T^{1,1}}$. 

\begin{figure}[h]
\begin{center}
\includegraphics[width=0.8\textwidth
]{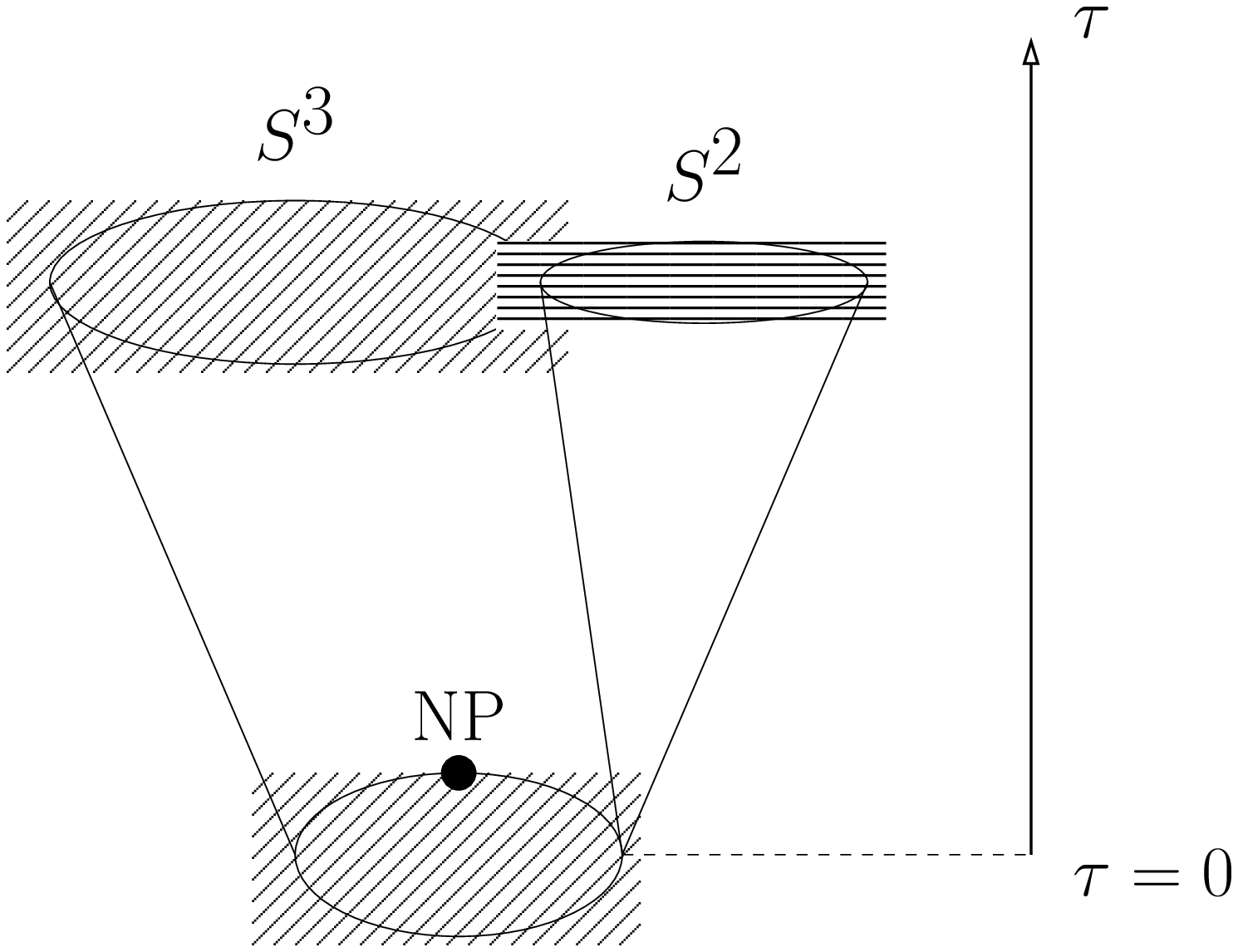}
\caption{A schematic picture of the (deformed) conifold. NP stands for the 
North Pole of the non-vanishing  three sphere at the tip. Our D-branes are 
at NP.
} 
\label{deformed}
\end{center}
\end{figure}

The singularity at the apex of the conifold can be smoothed in two ways 
while still respecting the Calabi-Yau condition as explained in the 
introduction. 
We will be studying the deformed case here, the resolved conifold has been 
subjected to a similar study in \cite{Klebanov:2007us}. For a
review of the various conifolds, see the appendices of 
\cite{Gwyn:2007qf}. A schematic picture of the conifold is in Figure 
\ref{deformed}.

The deformation of the conifold is defined by 
\eqn{cone1}{\sum_{i=1}^{4}z_i^2=\epsilon^2,}
which can be rewritten with an eye for useful future parametrizations as
\eqn{cone2}{{\rm det} W=-\frac{\epsilon^2}{2}, \ {\rm where} \  \ 
W\equiv \left(
\begin{array}{cc}
w_{11}   &  w_{12}   \\
w_{21} & w_{22}
\end{array}
\right)
=\frac{1}{\sqrt{2}}\left(
\begin{array}{cc}
z_3 + i z_4   &  z_1 - i z_2   \\
z_1 + i z_2 & -z_3 + i z_4
\end{array}
\right).}
By looking at the situation when all $z_i$ are real, it is clear that the 
$S^3$ does not vanish at the tip. Also worth noticing is the fact that the 
deformation breaks the $z_i \rightarrow e^{i\alpha} z_i$ symmetry of the 
singular conifold  down to $z_i \rightarrow -z_i$. So the deformed 
conifold does not have the full $U(1)$, but only a $\IZ_2$. The radius of 
the three sphere can be taken as 
\eqn{}{r^2 \equiv \sum_{i=1}^{4}|z_i|^2.}
It should be noted that this $r$ does {\em not} reduce to the radial 
coordinate of the cone in the undeformed limit. In fact, if we 
defined such a radial-like coordinate (i.e., a coordinate that 
tends to the radial coordinate of the undeformed conifold, far away 
from the deformation)  it would 
behave as $\tilde r \sim r^{2/3}$. We will use this information later.  

It is customary to use a new coordinate $\tau$ such that $r^2=\epsilon^2 
\cosh \tau$, in terms of which the above equation becomes
\eqn{tau}{{\rm Tr}(W^{\dagger}W)=\epsilon^2 \cosh \tau.}
The tip where the $S^2$ shrinks to zero corresponds to $\tau=0$.
	
Part of our purpose in the rest of this paper will be to use the metric 
to find the explicit supergravity solution that corresponds to a stack 
of D3-branes localized on the non-vanishing $S^3$. The D3-branes 
back-react and warp the geometry and we want to calculate the warp factor. 
To do this, we will need the Laplacian on the deformed conifold and it 
will be convenient to have a parametrization of $W$ where the split 
between the $S^2$ and the $S^3$ is explicit. 
The usual form in which the deformed conifold metric is written down 
does not have this advantage, so now we consider 
a system of coordinates where this split is manifest.


The aim is to package the information in the matrix $W$ into two 
separate pieces which can be interpreted as the $S^2$ and the $S^3$.
We start with the observation that the hermitian matrix $W^\dagger W$ has 
two real \emph{positive} eigenvalues:
\beq
\lambda_1^2=\frac{\epsilon^2 }{2} e^\tau \qquad
\textrm{and} \qquad
\lambda_2^2=\frac{\epsilon^2 }{2} e^{-\tau}.
\eeq
Taking positive square roots of $\lambda_1^2$ and $\lambda_2^2$ we can 
define a hermitian non-singular matrix $P \equiv \left( W^\dagger W 
\right)^{1/2}$ with the eigenvalues $\lambda_1,\lambda_2 > 0$.
This matrix, in turn, can be diagonalized:
\beq
P = U D(\tau) U^\dagger,
\qquad
\textrm{where}
\qquad
D(\tau)  \equiv
\left(
\begin{array}{cc}
 \lambda_1 & 0 \\
 0 & \lambda_2
\end{array}
\right)
\eeq
and $U$ is an $SU(2)$ matrix. Clearly $P$ is invariant under $U \to e^{i 
\alpha \sigma_3} U$ for any $\alpha$, so we have to quotient 
$U$ by this $U(1)$ action, which is 
just the Hopf projection from $SU(2)$ ($=S^3$)  to $S^2$. Thus $U$ 
describes the $S^2$. To 
build the $S^3$ we define a new  matrix $X$:
\beq
X \equiv -i W P^{-1}.
\eeq
It is easy to check that $X$ is unitary and special, so $X \in SU(2)=S^3$.

To summarize, for fixed $\tau$ we built a map from $W$ to $U$ and $X$ 
which defines the $S^2$ and the $S^3$ respectively. The map is invertible 
and simply given by:
\beq
\label{eq:WXP}
W=i XP=i X U D(\tau) U^\dagger.
\eeq
Moreover, for $\tau=0$ we find $P \propto \mathbb{I}_{2\times2}$ and so 
$U$ is ill-defined, which, as expected, means that for $\tau=0$ the 
two-sphere shrinks to zero size. Furthermore, for $\tau \to \infty$ we 
have
$P=r \left( \mathbb{I}_{2\times2} + iQ \right)$, where $Q$ is $2 \times 
2$ unitary anti-hermitian matrix and therefore the formula (\ref{eq:WXP}) 
re-produces the trivialization of the singular conifold proposed in 
\cite{EK}.

Now, we wish to write the deformed conifold metric not in terms of the 
original coordinates which mix the $S^2$ and $S^3$, but in terms of the 
coordinates that manifest the split. This is easily done because we just 
have 
to parametrize $U$ in terms of the angles of the two-sphere, and $X$ 
in terms of the angles of the three-sphere. We will follow 
the notations of \cite{Gimon:2002nr} and introduce two matrices $T$ and 
$S$ which are equivalent to our $X$ and $U$.\footnote{Notice that what we 
have provided 
essentially is an 
explicit construction of the $S$ and $T$ matrices of \cite{Gimon:2002nr} 
in terms 
of the standard conifold coordinates, captured by $W$.} We have:
\beq
X = -i T \sigma_3 \qquad
\textrm{and}   \qquad
U = \sigma_3 S \sigma_3, \quad
\textrm{where} \quad
S=e^{\frac{i}{2} \phi \sigma_3} e^{-\frac{i}{2} \theta \sigma_2}.\label{S}
\eeq 
This last bit defines a specific angular parametrization on the $S^2$ in 
terms of $\theta$ and $\phi$.  Once we also make a parametrization of the 
$SU(2)$ matrix $T$ in terms of the three angles of $S^3$ (which we 
write down in Appendix A), we will be done, and have explicit coordinates 
on the deformed conifold in terms of $\tau$, the three-sphere angles, and 
the two-sphere angles. Moreover, since it is well-known how to write $W$ 
(and therefore $X$ and $U$) in terms of the standard 
Klebanov-Strassler coordinates, we also have an explicit transformation 
relating the two coordinate systems.

To write the metric in a convenient form, we use the 
Maurer-Cartan forms $w_{i=1,2,3}$ on the three-sphere, defined by
\beq
T^\dagger \d T = \frac{i}{2} \sigma_i w_i.
\eeq
In terms of these angle coordinates and using 
(\ref{metricdeform}), the deformed conifold metric takes 
the following form:
\bea
\label{eq:metric6}
\epsilon^{-4/3} \d s_{(6)}^2 &=& \frac{1}{6 K^2(\tau)} \left( \d\tau^2 + 
h_3^2 
\right) +
                        \frac{K(\tau)}{4} \cosh^2 \left( \frac{\tau}{2} 
\right) \biggl[ h_1^2 + h_2^2 +  \\ \non
                      &&  \qquad  \qquad
   + 4 \tanh^2 \left( \frac{\tau}{2} \right) \left( (\d \theta -\half h_2)^2 + 
        (\sin \theta \d \phi-\half h_1)^2 \right)  \biggr].
\eea
Here

\beq
K(\tau) = \frac{\left( \sinh (2\tau) - 2\tau \right)^{1/3}}{2^{1/3} \sinh 
(\tau)}
\eeq
and the forms $h_{i=1,2,3}$ are defined by\footnote
{These forms are related to the analogous forms used in 
\cite{Gimon:2002nr} as follows:
$h_1=\sqrt{2} \tilde{g}^3$, $h_2=\sqrt{2} \tilde{g}^4$ and
$h_3= \tilde{g}^5$.
Notice also that $\sum_{i=1}^3 h_i^2=\sum_{i=1}^3 w_i^2$.}:
\beq
\label{eq:hw}
\left( \begin{array}{c} h_1 \\ h_2 \\ h_3 \end{array} \right) =
\left( \begin{array}{ccc} 0 & \cos \theta & -\sin \theta \\ 1 & 0 & 0 \\ 
0 & \sin \theta & \cos \theta \end{array} \right)
\left( \begin{array}{ccc} \sin \phi & \cos \phi & 0 \\ \cos \phi & -\sin 
\phi & 0 \\ 0 & 0 & 1\end{array} \right)
\left( \begin{array}{c} w_1 \\ w_2 \\ w_3 \end{array} \right).
\eeq
The two $SO(3)$ matrices in (\ref{eq:hw}) reflect the fact that the 
three-sphere is fibered over the two-sphere. This fiber is 
trivial as one can easily verify by properly calculating the Chern class 
of the fiber bundle \cite{EK}. We explicitly write down the $h_i$ in 
Appendix A in terms of the angles of $S^3$. 

From the metric (\ref{eq:metric6}) it is clear that at $\tau=0$ the size 
of the $S^2$ parameterized by $\theta$ and $\phi$ smoothly
shrinks to zero\footnote{Note that $K(\tau=0)=\left( \frac{2}{3} \right)^{1/3}.$}:
\beq
\epsilon^{-\frac{4}{3}} \d s_{(6)}^2 \approx 
\frac{1}{4} \left( \frac{2}{3} \right)^{\frac{1}{3}}
   \left[ \sum_{i=1}^3 w_i^2 +  \d \tau^2  + \tau^2
              \left( \left(\d \theta -\frac{h_2}{2} \right)^2 
  + \left(\sin \theta \d \phi-\frac{h_1}{2} \right)^2 \right) \right].
\eeq

\section{\bf D3-Brane Supergravity}
\label{LE}

The type IIB supergravity solution with D3-brane sources is fully
specified once we solve the Poisson-type equation for the warp factor
on the 6d space. We are interested in putting the stack of branes
at the deformed tip, where the $S^2$ has collapsed to zero size. This means
that we can look for the warp factor which is independent of $\theta$
and $\phi$. On top of that, without loss of generality, we will put the
D-branes at the North pole of the $S^3$, so that only the angle 
$\alpha$ (see Appendix A) will make its appearance in the warp factor. 
Arguments entirely
analogous to this were made in \cite{KK} in a different context, where a 
more detailed discussion
can be found. Using the Laplacian written down in Appendix B, and the 
above-mentioned simplifications, the final 
form of the warp-factor equation that we need to solve is
\beq
\Box_{\tau}H+\frac{1}{A^2(\tau)}\frac{1}{\sin^2 
\alpha}\partial_\alpha(\sin^2 \alpha \ \partial_\alpha H)=-\frac{6 
C}{\pi^2 \epsilon^4 \sinh^2 \tau \sin^2 
\alpha}\delta(\tau-\tau_0)\delta(\alpha). \label{warpeqn}
\eeq
The stack is at $\tau_0=0$.
$\Box_{\tau}$ and $A(\tau)$ are defined in Appendix B.
The general strategy for fixing the normalization of such delta 
functions and solving equations of this kind can be found 
in \cite{KK}. Here, $C=(2\pi)^4 g_s N \alpha'^2$, $N$ is the number of 
D3-branes. It is useful also to notice that the determinant of the 6d 
metric is
\beq
\sqrt{g_6}=\frac{\epsilon^4}{96}\sinh^2 \tau \sin^2 \alpha \sin \beta \sin 
\theta,
\eeq
where $\alpha$ and $\beta$ are the first two angles of the $S^3$ 
and
$\theta$ is the first angle (the latitude) of the $S^2$ (See 
Appendix A).

We first solve the angle part and look for solutions of 
\beq
\frac{1}{\sin^2
\alpha}\partial_\alpha(\sin^2 \alpha \ \partial_\alpha Y_l)+l(l+2) Y_l=0.
\eeq
We have chosen this form because energy eigenvalues of 
the $d$-sphere are of the form $l(l+d-1)$. The solutions of this 
three-sphere equation are in fact {\em simpler} than those of the familiar 
two-sphere, where the $Y_l$ take the well-known Legendre form. Here 
instead, we can take the independent solutions in the form
\beq
Y_l(\alpha)\sim \frac{\cos \left( (l+1)\alpha \right) }{\sin \alpha}, 
                \ \ \frac{\sin \left( (l+1)\alpha \right) }{\sin \alpha}.
\eeq
Of the two, since the right hand side of (\ref{warpeqn}) is even under 
$\alpha \leftrightarrow -\alpha$, we will only need the second set to do 
our expansions. We can fix the normalization by setting 
\beq
\int_0^{\pi} Y_l(\alpha) Y_{l'}(\alpha) \sin^2 \alpha \ d\alpha= 
\delta_{l l'}.
\eeq
The weight comes from the normalization of the delta function in the 
warp factor equation above. This fixes
\beq
Y_l(\alpha)=\sqrt{\frac{2}{\pi}}\frac{\sin \left( (l+1)\alpha \right) }{\sin \alpha}.
\eeq

Now, we turn to the radial equation, which takes the formidable shape
\beq
\Box_\tau H_l- \frac{l(l+2)}{A^2(\tau)}H_l(\tau)=-\frac{6C}{\pi^2 
\epsilon^4 \sinh^2 \tau}\delta(\tau-\tau_0). \label{radial}
\eeq
We have been able to solve this equation for generic $l$ only 
numerically\footnote{For $l=0$, there is a slight simplification. The 
solution can be written as
\beq
H_{l=0}(\tau) \sim \int^{\tau}\frac{1}{(\sinh 2x -2x)^{2/3}}
{\rm d}x.
\eeq
}. 
To 
fully fix a second order differential equation, we need two pieces of 
data (e.g.: the value of the function at two different points or the value 
of the function and its derivative at the same point.). The homogeneous 
equation only determines the solution upto an overall constant, even 
after one stipulates that it die down at infinity. This 
overall normalization is fixed by the strength of the delta-function 
discontinuity at the origin. In particular, in our case it turns out that 
this gives, 
\beq
\lim_{\tau \rightarrow 0}\biggl[H_l'(\tau)\left( \sinh 
(2\tau)-2\tau\right)^{2/3}\biggr] = -\frac{ 
2^{2/3}}{\pi^2\epsilon^{8/3}} C.
\eeq
So one numerical consistency check we can do on our solutions is to check 
that the 
left hand side has a good limit as $\tau \rightarrow 0$.

\begin{figure}[h]
\begin{center}
\includegraphics[width=0.8\textwidth
]{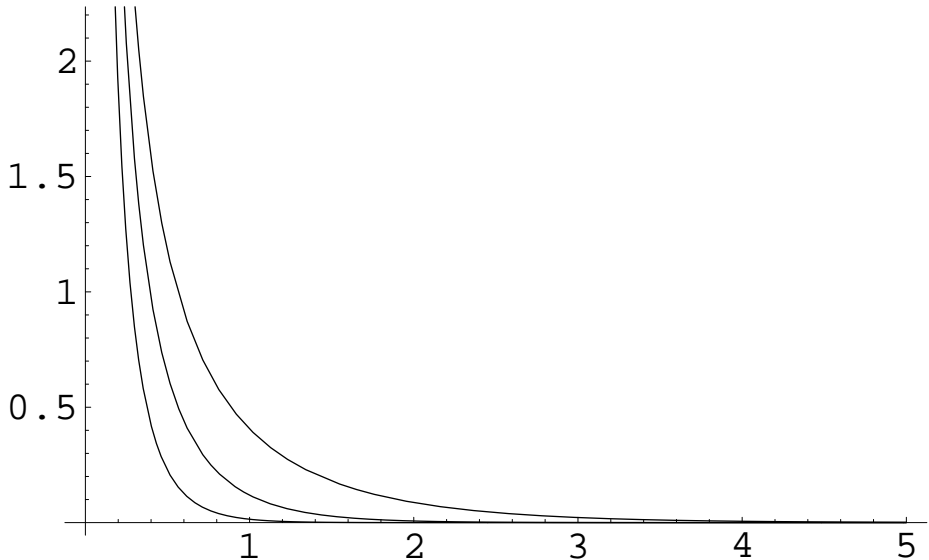}
\caption{Plots of $\alpha H_l(\tau)$ where $\frac{1}{\alpha}=\frac{
2^{2/3}}{\pi^2\epsilon^{8/3}} C$, for $l=0, 1, 3$. The curve rises as $l$ 
increases.
}
\label{plot}
\end{center}
\end{figure}

We can do another check. We can solve the asymptotic ($\tau 
\rightarrow \infty$) form of the differential equation {\em exactly}. The 
asymptotic (homogeneous) equation takes the form
\beq
h_l''(\tau)+\frac{4}{3}h_l'(\tau)-\frac{4}{3}l(l+2)h_l(\tau)=0.
\eeq
The dying solutions of this equation are
\beq
h_l(\tau) \sim \exp \bigg[- 
\frac{2\tau}{3}\big(1+\sqrt{1+6l+3l^2}\big)\bigg],
\eeq
which implies that 
\beq
\lim_{\tau \rightarrow \infty} 
\frac{H_l'(\tau)}{H_l(\tau)}=-\frac{2}{3}\big(1+\sqrt{1+6l+3l^2}\big).
\eeq
This is easily checked numerically, and indeed we have checked that it is 
satisfied for our solutions. A plot of the radial solutions for some 
values of $l$ are given in Figure \ref{plot}.

The full solution then, can be written as
\bea
H(\tau,\alpha)=\sum_{l=0}^{\infty} H_l(\tau)\ Y_l(\alpha)\ 
Y_l(\alpha_0=0)= 
\frac{2}{\pi}\sum_{l=0}^{\infty}(l+1) H_l(\tau) \frac{\sin \left((l+1)\alpha\right) }{\sin \alpha}. \label{full}
\eea

One rather basic consistency check that we can do with this full 
solution is to compare it to the smeared 
approximation: the $l=0$ term of the above sum should reproduce the 
results obtained by 
assuming that the D3-branes were smeared on the 3-sphere. When the branes 
are smeared, the warp factor equation is the Laplace equation with 
all angular dependence suppressed, which reduces to
$\Box_\tau H = 0$. The overall normalization can be fixed either by 
comparison with undeformed conifold in the asymptotic region, or by 
being careful about the normalization of the delta function source. This 
normalization essentially just amounts to an extra factor of 
$\frac{2}{\pi}$ from the integration of the $\sin^2 \alpha$ that was there 
in the delta function before the smearing. But from (\ref{full}), we see 
that it is precisely a factor of $\frac{2}{\pi}$ that multiplies the 
$H_l(\tau)$, when $l=0$. 

Using the warp factor, one can also demonstrate the emergence of the AdS 
throat close to the D-brane stack. This is easiest to do along 
$\alpha=0$\footnote{We can expand the deformed conifold metric from the 
previous section when $\tau, \alpha \ll 1$. The radial coordinate turns 
out to be of the form $\sim \sqrt{\tau^2 +\alpha^2}$ upto irrelevant 
numerical factors. By restricting to a flow along which
$\alpha=0$, our radial coordinate takes the simpler form $\tau$.}. There 
the warp 
factor (\ref{full}) takes the form $\sim 
\sum (l+1)^2H_l(\tau)$. Now, we can take a ``near-horizon" limit of  
(\ref{radial}) and solve it exactly to find what $H_l(\tau)$ looks like 
close to the stack. It turns out that the (homogeneous) near-horizon 
radial equation is
\beq
H''(\tau)+\frac{2 H'(\tau)}{\tau}-l(l+2) H(\tau)=0, \ {\rm with \ 
solution}
\ H(\tau) \sim \frac{e^{-\sqrt{l(l+2)}\tau}}{\tau}.
\eeq
It turns out that the normalization of the solution (fixed by 
integrating across the source) is independent of $l$, so the 
{\em entire} dependence on $l$ and $\tau$ is captured by the above 
expression, which we write schematically as $\frac{f(l\tau)}{\tau}$. Since 
the sum over all $l$'s must converge, we 
can think of this as a regulator \cite{Klebanov:2007us} and 
write
\beq
H(r)\sim \sum_{l=0}^{\infty}l^2 \frac{f(l 
\tau)}{\tau}\sim\sum_{l=0}^{1/\tau}l^2 \frac{f(l \tau)}{\tau}
\sim \int_0^{1/\tau}l^2 \frac{f(l\tau)}{\tau} dl \sim \frac{\int_0^1x^2 
f(x) dx}{\tau^4}\sim\frac{\rm const.}{\tau^4}.
\eeq  
Since the radial coordinate looks like $\tau$ near $\tau\sim0$ (see 
footnote), this means that in the near horizon region, in terms 
of the flat coordinate, the warp factor goes as $\sim \frac{1}{\tau^4}$. 
But this is of course what gives rise to the origin of the AdS throat.

\section{\bf The Dual Gauge Theory and the Mesonic Branch}
\label{GT}

As we have explained in the Introduction our supergravity solution 
describes a stuck of $N$ $D3$-branes located at the ``tip" of the deformed conifold. The 
gauge theory dual to this solution was analyzed both in the original 
paper \cite{KS} and in more detail in \cite{DKS}.
The dual theory has an $SU(\widetilde{N}+M) \times SU(\widetilde{N})$  
gauge group, where $\widetilde{N}$ is related to $M$ and $N$ as in (\ref{eq:tildeN}). As 
the $SU(\widetilde{N}+M)$ gauge group becomes strongly coupled in the 
IR, it is described effectively by four mesons $M_{\alpha  
\beta}=A_{\alpha} B_{\beta}$, where $A_{1,2}$ and $ B_{1,2}$ are the 
bi-fundamental chiral fields. The theory is Seiberg dual to a theory with 
an $SU(\widetilde{N}) \times  SU(\widetilde{N}-M)$ gauge group, where now 
the first factor becomes strongly coupled in the IR, and the field 
content is given by the dual ``magnetic" quarks, which play now the role 
of the bi-fundamental fields. For each step of the cascade, therefore, we 
have  $\widetilde{N} \to \widetilde{N} - M $.  In general, the duality 
cascade proceeds until $\widetilde{N}$ becomes smaller than $M$, where the 
dual description does not exist and instead the quantum corrections are 
captured by the non-perturbative Affleck-Dine-Seiberg (ADS) term in the 
superpotential. We are, however, interested in a case where the cascade 
stops at $\widetilde{N}=N$, due to the mesons acquiring
large enough VEVs. In this situation the classical superpotential also 
receives a non-perturbative ADS like contribution:
\beq
\label{eq:W}
W = h \Tr \left( \CM_{11} \CM_{22} - \CM_{12} \CM_{21} \right) + 
    (M-N) \left( \frac{\Lambda^{N+3M}}{ \det_{ab \alpha \beta} \CM} 
\right)^{\frac{1}{M-N}} .
\eeq
Here the first term is the classical superpotential.
Notice that for $N>M$ the determinant appears actually with a positive 
power. It becomes a real ADS potential only for $N<M$.  In any case, 
however, the moduli space describes $N$ $D3$-branes moving on the 
deformed conifold.
Indeed, the equations of motion for the mesons imply that all the 
matrices $M_{\alpha \beta}$ commute and also:
\bea
\det_{ab\alpha \beta} \CM &=&  \left( 
                 h^{ \left(\frac{N}{M}-1 \right)} \Lambda^{\left(\frac{N}{M}+ 3 \right)}  \right)^{N}
\non \\
\Tr_{ab} \det_{\alpha \beta}  \CM &=& 
                 N h^{ \left(\frac{N}{M}-1 \right)} \Lambda^{
\left(\frac{N}{M}+3\right)}.
\eea
The matrices  $M_{\alpha \beta}$ can be all simultaneousely 
diagonalized. The above equations will then both 
lead to the deformed conifold definition for the eigenvalues 
of $M_{\alpha \beta}$'s with the deformation parameter $\epsilon$ being 
a function of $h$ and $\Lambda$:
\beq
\epsilon^2 \propto  h^{ \left(\frac{N}{M}-1 \right)} \Lambda^{\left(
\frac{N}{M}+3\right)}.
\eeq
Computations similar to what we have done above can be found, for example, 
in \cite{Berenstein:2001uv, Berenstein:2002sn}. In the latter, matrix 
model techniques were applied for steups with more than one conifold 
singularity.

In this paper we have constructed a supergravity solution dual to the 
mesonic branch of the gauge theory.
The $D3$-brane  source in our picture is located at $\tau=0$ (the 
minimal value of the radial cootdinate),
where the two sphere smoothly shrinks to zero. For the $10d$ solution 
to be regular the $D3$-branes have to be localized at a point, which 
in our conventions  is the North pole of the non-shrinking three-sphere.
On the gauge theory side it means that all the eigenvalues 
$m^i_{\alpha \beta}$ of the matrix $\CM_{\alpha \beta}$ are the same 
and correspond to the North pole of the $S^3$
as we have explained in Section \ref{DCP}. 
An RG flow triggered by the VEVs leads in the IR to the $\mathcal{N}=4$ 
SYM theory, which on the supergravity side
corresponds to $AdS$ throat developed near the $D3$-brane source.
In the rest of this Section we want to show that expanding the 
superpotential (\ref{eq:W}) around the VEV corresponding to the North 
pole we find, as expected, the cubic 
superpotential of  the $\mathcal{N}=4$ SYM.

In the coordinates introduced in Appendix A the North pole
corresponds to $\alpha=0$ and thus $X=\sigma_0$.  On the other hand $D(\tau=0)=\frac{\epsilon}{\sqrt{2}} \sigma_0$
and so (\ref{eq:WXP}) implies that $W=i\frac{\epsilon}{\sqrt{2}} \sigma_0$.
We know that $m^i_{\alpha \beta}$'s are related to $w_{\alpha \beta}$'s so
we have to consider the following VEVs:
\beq
\left< \CM_{11} \right> = i\frac{\epsilon}{\sqrt{2}} \cdot \mathbb{I}_{N \times N}, \quad
\left< \CM_{22} \right> = i\frac{\epsilon}{\sqrt{2}} \cdot \mathbb{I}_{N \times N}, \quad
\left< \CM_{12} \right> = 0, \quad
\textrm{and} \quad
\left< \CM_{21} \right> = 0.
\eeq
Next we will consider the expansion around the VEV:
\bea
\CM_{11}  = \left< \CM_{11} \right> + \delta \cdot \left( \Phi - 
\Phi_1 \right)  &,& \quad
\CM_{22}  = \left< \CM_{22} \right> + \delta \cdot \left( \Phi + 
\Phi_1 \right),   \non \\
\CM_{12}  = \left< \CM_{12} \right> + \delta \cdot \Phi_2 &,& \quad
\CM_{12}  = \left< \CM_{21} \right> + \delta \cdot \Phi_3,
\eea
where 
\beq
\delta^2 \equiv  h^{ \left(-1 + \frac{N}{3M} \right)} \Lambda^{ 
\left(1 + \frac{N}{3M}\right)}.
\eeq
Up to the quartic terms this yields:
\beq
W = \textrm{const} + \Tr \left( \Phi_1 \left[ \Phi_2, \Phi_3 
\right] \right) +
    2 h \delta^2 \Tr \Phi^2 + \frac{2}{3} \Tr \Phi^3 + \Tr \left( 
\Phi \left\{ \Phi_2, \Phi_3 \right\} \right) + 2 \Tr \left(\Phi 
\Phi_1^2\right)
  \ldots
\eeq
Here we made use of the formulae collected in Appendix C.
The first non-trivial term here is exactly the $\mathcal{N}=4$ SYM 
cubic superpotential. Notice that
the fields $\Phi_i$ have dimension one as it should be in the 
conformal $\mathcal{N}=4$ theory. It follows
from the fact that $\CM_{\alpha \beta}$ have dimension two and 
the parameter $\delta$ has dimension one.
The remaining field $\Phi$  is massive. This is expected, since 
the deformed conifold is a three-dimensional embedding in 
$\mathbb{C}^4$  and $\Phi$ describes the only direction, which 
is not tangent to the conifold.
Thus this field is also not tangent to the moduli space and is 
expected to be massive.
One can easily check that integrating out $\Phi$  produces 
quartic $\Phi_i$ terms, which, of course,
become irrelevant in the IR.

\section{\bf Concluding Remarks}
\label{D}

In this paper we have constructed  a supergravity background 
dual to the mesonic branch of the gauge theory. We therefore do not expect the $U(1)_B$ to be 
broken. 
The baryonic symmetry is not related to one of the background isometries,
it rather appears as a gauge symmetry of the Wess-Zumino term.
Still one can ask whether the Goldstone boson mode 
found in \cite{Gubser}
ceases to exist once we add the $D3$-brane source.  
If the mode does not exist anymore we can safely assume that the 
baryonic symmetry is unbroken.
This indeed seems to be the case,
since most of the expressions (for example equation (3.25))  in 
\cite{Gubser} explicitly include the warp 
function $h$. For $h=h_{\textrm{KS}}$ these 
expressions are normalizable at $\tau \to 0$ 
with respect to the conifold metric. However, for $h=h_{\textrm{KS}}+H_{\textrm{D3}}$
most of these expressions will diverge, since near the North pole at $\tau=0$ we have 
$h \approx H_{\textrm{D3}} \approx\frac{1}{\tau^4}$.  It will 
be interesting to make this statement more rigorous
proving therefore that our background is indeed related to the mesonic branch of the gauge theory, where
no Goldstone boson is expected.

\section{Acknowledgments}

It is a pleasure to thank Riccardo Argurio, Cyril Closset, Jarah 
Evslin and Carlo Maccaferri for useful 
conversations.
This work is supported in part by IISN - Belgium (convention 
4.4505.86), by the Belgian National
Lottery, by the
European Commission FP6 RTN programme MRTN-CT-2004-005104 in which the 
authors
are associated with V. U. Brussel, and by the Belgian Federal Science
Policy Office through the Interuniversity Attraction Pole P5/27.

\section*{\bf Appendix}

\addcontentsline{toc}{section}{Appendix}

\subsection*{{\bf A} \ Technicalities: $S^2$, $S^3$ and the K\"ahler 
Metric}
\addcontentsline{toc}{subsection}{{\bf A} \ Technicalities: $S^2, S^3$ 
and the K\"ahler Metric}
\renewcommand{\theequation}{A.\arabic{equation}}

We complete the definition of $S^2$ and $S^3$ here by giving the 
explicit matrices. Using the matrix $S$ defined in (\ref{S}) the matrix 
$P$ becomes:
\eqn{}{
P=\frac{\epsilon}{\sqrt{2}}\left(
\begin{array}{cc}
\ e^{\frac{\tau}{2}}\cos^2 \frac{\theta}{2}+\ 
e^{-\frac{\tau}{2}}\sin^2\frac{\theta}{2}   
&  \sinh (\frac{\tau}{2})e^{i\phi}\sin \theta    \\
\\
\sinh (\frac{\tau}{2})e^{-i\phi}\sin \theta 
 & \ 
e^{-\frac{\tau}{2}}\cos^2 \frac{\theta}{2}+\ 
e^{\frac{\tau}{2}}\sin^2\frac{\theta}{2}
\end{array}
\right).
}

As for the three-sphere $S^3$, it is defined by the real 
numbers 
satisfying
\beq
x_0^2+x_1^2+x_2^2+x_3^2=1. \label{3sphere}
\eeq
This is identical to the group $SU(2)$ because (\ref{3sphere}) is 
precisely the 
condition 
that turns a general 2 $\times $ 2 matrix $X$ defined by 
\eqn{}{
X=\left(
\begin{array}{cc}
x_0 + i x_3   &  i x_1 + x_2   \\
i x_1 - x_2 & x_0 - i x_3
\end{array}
\right)
}
into a special, unitary 2 $\times $ 2 matrix.
Using the fact that $SU(2)$ is a group, we can use its Maurer-Cartan 
one form
\eqn{xdaggerdX}{X^{\dagger}dX\equiv\frac{i}{2}\sum_{i=1}^{3}w_i\sigma_i,}
to define a basis of canonical one-forms on $S^3$. If we parametrize $S^3$ 
in the usual way
\bea
x_0= \cos \alpha, \ x_1= \sin \alpha \cos \beta, \ x_2= \sin \alpha \sin 
\beta \cos \gamma, \ x_3= \sin \alpha \sin \beta \sin \gamma, 
\eea
then, by explicit computation using the above formulae, we find
\begin{eqnarray}
\frac{w_1}{2}&=&\cos\beta 
d\alpha-\sin\alpha\cos\alpha\sin\beta 
d\beta+\sin^2\alpha\sin^2\beta 
d\gamma, 
\\ 
\frac{w_2}{2}&=&\sin\beta\cos\gamma d\alpha+(\sin\alpha\cos\alpha
\cos\beta\cos\gamma-\sin^2\alpha\sin\gamma) 
d\beta+ \nonumber \\
&&\hspace{0.6in}-(\sin^2\alpha\sin\beta\cos\beta\cos\gamma+
\sin\alpha\cos\alpha\sin\beta\sin\gamma 
)d\gamma, \\
\frac{w_3}{2}&=& \sin\beta\sin\gamma 
d\alpha+(\sin\alpha\cos\alpha\cos\beta\sin\gamma+\sin^2\alpha\cos\gamma 
)d\beta +\nonumber \\
&&\hspace{0.6in}+(\sin\alpha\cos\alpha\sin\beta\cos\gamma-
\sin^2\alpha\sin\beta\cos\beta\sin\gamma 
)d\gamma.
\end{eqnarray}

Now we turn to the conventional definition of the metric on the deformed 
conifold in terms of its K\"ahler potential. We use this in 
the derivation of (\ref{eq:metric6}). The metric can be written in the
form \cite{Candelas:1989js}:
\eqn{metricdeform}{ds_6^2=
{\cal F}' \  {\rm
Tr}({\rm d}W^{\dagger}{\rm d}W)+
{\cal F}'' \ |{\rm Tr}
(W^{\dagger}{\rm d} W)|^2,}
where ${\cal F}\equiv{\cal F}(r^2)$ and
\begin{eqnarray}
{\cal F}'\equiv\frac{\partial{\cal F}}{\partial r^2}=
\frac{1}{\epsilon^2}\frac{1}{\sinh \tau}\frac{\partial {\cal F}}{\partial
\tau},\ \  {\rm with}\  \ \frac{\partial {\cal F}}{\partial
\tau}=2^{-1/3}\epsilon^{4/3}(\sinh (2\tau)- 2 \tau)^{1/3}.
\end{eqnarray}

\subsection*{{\bf B} \ Laplacian in Two Different Coordinates}
\addcontentsline{toc}{subsection}{{\bf B} \  Laplacian in Two Different 
Coordinates}
\renewcommand{\theequation}{B.\arabic{equation}}

This appendix is dedicated to writing down the Laplacian for the deformed 
conifold in the standard coordinates and also in the coordinates where the 
$S^2-S^3$ split is manifest. We will need only the latter form, but we 
present both of them here for the convenience of posterity. In what 
follows, the functions $A(\tau)$ and $B(\tau)$ are defined by
\beq
A^2(\tau)=\frac{2^{-1/3}}{8}\coth \frac{\tau}{2}\ (\sinh 2\tau -2
\tau)^{1/3}, \ \
B^2(\tau)=\frac{2^{2/3}}{6}\frac{\sinh^2 \tau}{(\sinh 2\tau -2
\tau)^{2/3}}.
\eeq

\underline{\bf Klebanov-Strassler Coordinates:}

The scalar Laplacian in Klebanov-Strassler coordinates can be written 
in the form (the notations can be found in \cite{KS}):
\begin{eqnarray}
\Box H=\Box_\tau H+f_R(\tau)\Box_R H +
f_S(\tau)\big(\Box_1 H +\Box_2 H\big)+f_m(\tau)\Box_m H, \label{lapl}
\end{eqnarray}
where
\begin{eqnarray}
f_R(\tau)=\frac{1}{B^2(\tau)}, \ \ f_S(\tau)=\frac{\coth^2 
\tau}{A^2(\tau)} , \ \ f_m(\tau)=\frac{\cosh \tau}{A^2(\tau)\sinh^2\tau},
\end{eqnarray}
and
\begin{eqnarray}
\Box_\tau=\frac{\coth^2 \tau}{A^4(\tau)B^2(\tau)}\frac{\partial}{\partial 
\tau} \left(A^4(\tau)\tanh^2 \tau \frac{\partial}{\partial \tau} \
\right), \ \ \displaystyle{ \Box_R = \partial^2_\psi },
\end{eqnarray}
\begin{equation}
\displaystyle{
\Box_i = \frac{1}{\sin\theta_i} \partial_{\theta_i}  \;
               (\sin \theta_i \; \partial_{\theta_i} \;\;) +
               \left( \frac{1}{\sin\theta_i} \partial_{\phi_i} - 
\cot\theta_i
\partial_\psi \right)^2
}.
\end{equation}
The $\Box_i$ arise from the two $S^3$'s (or equivalently, $SU(2)$'s) that 
are part of the original $T^{1,1}=SU(2)\times SU(2)/ U(1)$. The modding by
the $U(1)$ is reflected in the fact that the two $S^3$ Laplacians share a 
common angle, $\psi$.
The ugly final piece in (\ref{lapl}) that could mix the various angular 
eigenvalues is:
\begin{eqnarray}
\frac{1}{2} \ \Box_m=-\cos \psi 
\left(\partial_{\theta_1}\partial_{\theta_2}-\Big( \cot \theta_1 
\partial_\psi -\frac{\partial_{\phi_1}}{\sin
\theta_1}\Big)\Big( \cot \theta_2 \partial_\psi 
-\frac{\partial_{\phi_2}}{\sin \theta_2}\Big)\right)+ \nonumber \\
+\sin \psi \left( \Big(\cot \theta_2 \partial_\psi -\frac{1}{\sin 
\theta_2} \partial_{\phi_2}
\Big)\partial_{\theta_1} +\Big(\cot \theta_1 \partial_\psi -\frac{1}{\sin 
\theta_1} \partial_{\phi_1}\Big)\partial_{\theta_2} \right). \label{mixed}
\end{eqnarray}

\underline{\bf $S^2-S^3$ coordinates:}

With $\Box_\tau, A(\tau)$ and $B(\tau)$ defined as in the previous case, 
we have:
\begin{eqnarray}
\Box H= \Box_\tau 
H+\frac{1}{A^2(\tau)}(\partial_1^2+\partial_2^2+\partial_3^2)H+\frac{1+\coth^2 
\frac{\tau}{2}}{4A^2(\tau)}\left(\frac{\partial_\theta(\sin 
\theta\partial_\theta 
H)}{\sin \theta}+\frac{\partial_\phi^2 H}{\sin^2\theta}\right)+ 
\hspace{-0.1in}\nonumber 
\\
+\frac{1}{A^2(\tau)}\bigg(\big(\sin \phi \partial_1+\cos \phi 
\partial_2\big)\partial_\theta H+\big(\cos \theta(\cos \phi 
\partial_1-\sin \phi\partial_2)-\sin \phi 
\partial_3\big)\frac{\partial_\phi H}{\sin \theta}\bigg). \nonumber \\
\end{eqnarray}
Here $\partial_i\equiv \partial_{w_i}, \ i=1,2,3.$. In particular, 
$(\partial_1^2+\partial_2^2+\partial_3^2)$ is nothing but the $S^3$ 
Laplacian. When we put the stack of D3-branes on the non-vanishing $S^3$, 
the $S^2$ has shrunk to zero size and so we can drop terms that have 
derivatives of $H$ with respect to $\theta$ and $\phi$.

\subsection*{{\bf C} \ Matrix Technology}
\addcontentsline{toc}{subsection}{{\bf C} \ Matrix Technology}
\renewcommand{\theequation}{C.\arabic{equation}}

Let us use the notation $\dCM = \det \CM$, where $\CM$ is an arbitrary 
\emph{invertible}
square matrix. Then:
\bea
\frac{\delta \dCM^n}{\delta \CM_{mn}} \delta \CM_{mn} &=& n \dCM^n 
\Tr (\CM^{-1} \delta \CM)     \non \\
\frac{\delta^2 \dCM^n}{\delta \CM_{mn} \delta \CM_{m^\prime n^\prime}} 
\delta \CM_{mn} \CM_{m^\prime n^\prime}          &=& n \dCM^n \left( n 
\left(\Tr (\CM^{-1} \delta \CM) \right)^2 -  \Tr \left(\CM^{-1} \delta 
\CM \right)^2 \right)   \non \\  
\eea 
and
\bea
&&
\frac{\delta^3 \dCM^n}
{\delta \CM_{mn} \delta \CM_{m^\prime n^\prime} \delta 
\CM_{m^{\prime\prime} n^{\prime\prime}}}  
         \delta \CM_{mn} \CM_{m^\prime n^\prime}  \delta 
\CM_{m^{\prime\prime} n^{\prime\prime}}   =
n \dCM^n \Big( n^2 \left(\Tr (\CM^{-1} \delta \CM) \right)^3 -
\non \\
&&
\qquad
- 3n \left(\Tr (\CM^{-1} \delta \CM) \right)
            \Tr \left(\CM^{-1} \delta \CM \right)^2  +2 \Tr 
\left(\CM^{-1} \delta \CM \right)^3   \Big).
\eea

\newpage

\end{document}